\documentclass[showkeys,showpacs,preprintnumbers,groupedaddress,amsmath,amssymb,floatfix]{revtex4}
\usepackage{graphicx}
\usepackage{epsfig}
\usepackage{ulem}

\begin{document}

\title{Using Charged Particle Imaging to Study Ultracold Plasma Expansion}

\author{X. L. Zhang, R. S. Fletcher, and S. L. Rolston}
\affiliation{Joint Quantum Institute, Department of Physics, University of Maryland, College Park, MD 20742}


\begin{abstract}

  We develop a projection imaging technique to study ultracold plasma dynamics. We image the charged particle spatial distributions by extraction with a high-voltage pulse onto a position-sensitive detector. Measuring the 2D width of the ion image at later times (the ion image size in the first 20 $\mu$s is dominated by the Coulomb explosion of the dense ion cloud), we extract the plasma expansion velocity. These velocities at different initial electron temperatures match earlier results obtained by measuring the plasma oscillation frequency. The electron image size slowly decreases during the plasma lifetime because of the strong Coulomb force of the ion cloud on the electrons, electron loss and Coulomb explosion effects. 
\end{abstract}

\maketitle
  Ultracold plasmas (UCPs), formed by photoionizing laser-cooled atoms near the ionization limit, have well-controlled initial conditions and slow dynamics compared to other laser-produced plasma systems, and thus provide a clean and simple source with excellent spatial and temporal resolution available to study basic plasma phenomena. In the majority of experiments to date, UCPs have been unconfined and freely expanded into vacuum, a fundamentally important dynamic in laser-produced plasmas as well as UCPs. The first experimental study of the expansion of UCPs was performed using the plasma frequency as a probe of the plasma density as a function of time \cite{kulin2000}. By applying a small RF electric field to excite plasma oscillations, the plasma density time dependence was mapped from the oscillations, and a ballistic expansion of the plasma was found, i.e. $\sigma^2(t) = {\sigma^2_0+v^2_0t^2}$. For initial electron temperatures $T_e(0)\geq$ 70 K, the expansion velocities follow $v_0^2 = k_BT_e/m_i$, the ion acoustic velocity due to the electron pressure on the ions, in agreement with a simple hydrodynamics model. At low initial temperatures, the UCPs expand faster than expected, which indicates plasma heating. The expansion dynamics of UCPs have subsequently been studied experimentally by various methods, such as plasma collective modes \cite{fletcher2006}, absorption imaging \cite{simien2004}, fluorescence imaging \cite{cummings2005}, and theoretically \cite{bergeson2003, robicheaux2003, pohl2004, mazevet2002}.

In this work we use a projection imaging technique to study the UCP dynamics during the full lifetime of the plasma. We image the charged particle (ions or electrons) spatial distribution of an expanding UCP by extracting them with a high-voltage pulse and accelerating them onto a position-sensitive detector. The expansion is self-similar, as the ion (or electron) cloud maintains a Gaussian density profile throughout the lifetime of the plasma. Early in the lifetime of the plasma, the ion image size is dominated by the Coulomb explosion of the dense ion cloud. The image size is at a minimum at about 20 $\mu$s and then afterwards increases (the Coulomb explosion of the ion cloud becomes negligible), reflecting the true size of the plasma. We obtain the ion image width by 2D Gaussian fitting, and extract the final asymptotic expansion velocity by fitting the linear region of the ion images as a function of time at later plasma times ($\geq$ 20 $\mu$s). Assuming that the ion cloud maintains the Gaussian density distribution during the Coulomb explosion phase, we can extract the actual ion cloud size from the ion projection image by accounting for the Coulumb explosion effect. The plasma size indeed follows the ballistic expansion as expected from a simple hydrodymics model throughout the whole lifetime of the plasma. Including the corrected plasma sizes in the first 20 $\mu$s compared to that obtained by only fitting the linear region at later times only amounts to a few percent change in asymptotic velocity. The velocities at different initial electron temperatures matches earlier results obtained by measuring the plasma oscillation frequency \cite{kulin2000}, which provides strong support for this method to study the UCP expansion and the previous technique.  We can also image the electrons during the lifetime of an UCP by switching the polarity of the high-voltage pulse and the accelerating voltages on the grids. Unlike the ion images, The size of the electron image slowly decreases during the lifetime of UCP because of the strong Coulomb effect of the dense ion cloud on the electrons, electron loss and Coulomb explosion effect. This technique provides a good tool to study UCP dynamics in a magnetic field, such as expansion \cite{zhang20081} and plasma instabilities \cite{zhang20082}. 

\begin{figure}[htbp]
\begin{center}
\epsfig{file=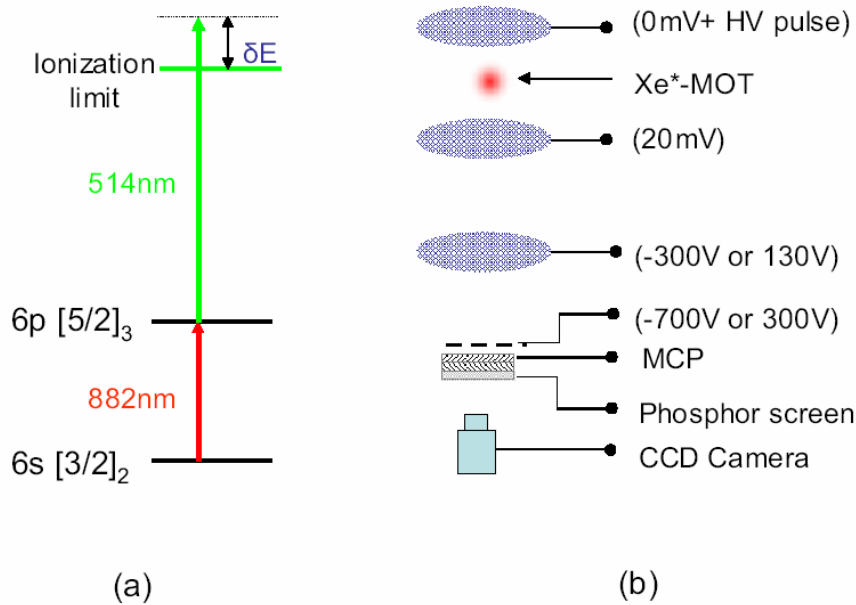, width=5in}
\end{center}
\caption{ Two-photon excitation process and experimental setup. (a) two-photon excitation process: one photon (red solid line) at 882 nm drives the $6s [3/2]_2$ $\rightarrow$  $6p [5/2]_3$ transition, and the other (10-ns pulse) at 514 nm ionizes the atoms in the $6p [5/2]_3$ state. (b) experimental setup for imaging the charged particles onto the MCP/phosphor screen.}
\end{figure}

Details of the creation of ultracold neutral plasmas are described in \cite{killian1999}. We accumulate about $10^6$ metastable Xenon atoms, trapped and cooled in a magneto-optical trap to a temperature of $\sim$ 20 $\mu$K. The atomic cloud has a Gaussian spatial density distribution with a peak density of about 2 x 10$^{10}$ cm$^{-3}$ and an rms-radius of $\sim$ 0.3 mm. We produce the plasma with a two-photon excitation process (882-nm photon from the cooling laser and 514-nm photon from a 10-ns pulsed dye laser), ionizing up to 30$\%$ of the atoms. We control the ionization fraction with the intensity of the photoionization laser, while the initial electron energy is controlled by tuning the 514-nm photon energy with respect to the ionization limit, usually in the 1-1000 K range. The ionized cloud rapidly loses a few percent of the electrons, resulting in a slightly attractive potential for the remaining electrons, and quickly reaches a quasineutral plasma state. It then freely expands with an asymptotic velocity $v_0$ typically in the 50-100 m/s range caused by the outward electron pressure\cite{kulin2000}.

\begin{figure}[htbp]
\begin{center}
\epsfig{file=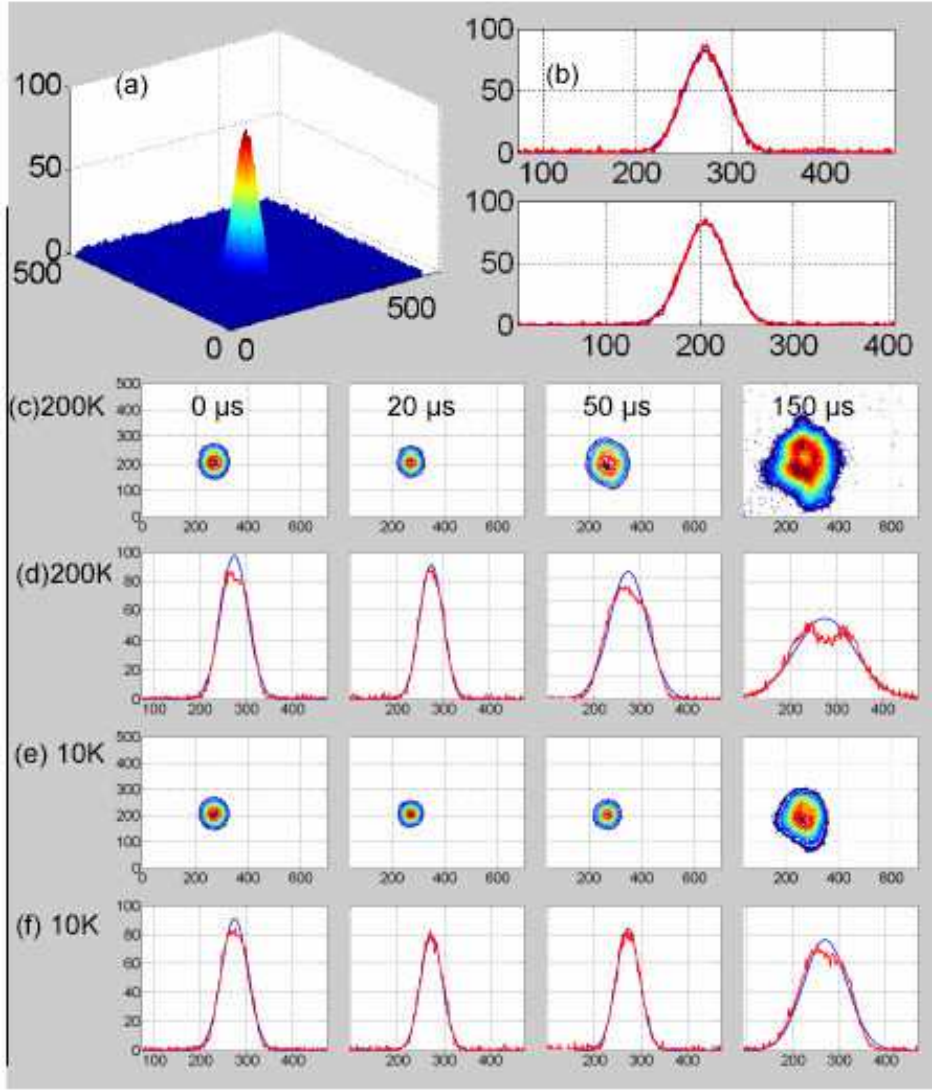, width=5in}
\end{center}
\caption{ (a) a false color ion image (2D ion spatial distribution integrated over the third dimension) of an expanding UCP at t = 20 $\mu$s, $T_e(0)$ = 100 K; (b) the 2-D Gaussian fittings (blue curves) of the ion image (a) along the x and y axis in the horizontal plane (red curves); (c)and (e) are the contour plots of the ion images at different delay times for $T_e(0)$ = 200 K and 10 K, respectively; (d) and (f) are the corresponding 2-D Gaussian fitings of (c) and (e). All the size related units in (a)-(f) are in pixel number, and one pixel unit corresponds to 150 $\mu$m. The y axis of (d) and (f) is in arbitrary unit.}
\end{figure}
 
For projection imaging of charged particles, external electric fields are applied via four grids to direct and accelerate them towards a position-sensitive detector (a micro-channel plate detector with phosphor screen) (figure 1b). Two grids ("top" and "bottom" grids) are 1.5 cm above and below the plasma, and the other two ("middle" and "front" grids) are located between the bottom grid and the detector. By applying a high-voltage pulse to the top grid at a specific delay time after the formation of the UCP and with accelerating voltages on the middle and front grids (-300 V and -700 V for ions, 130 V and 300 V for electrons), we image the charged particle distribution of expanding UCPs onto the phosphor screen. The phosphor image, recorded by a CCD camera, is proportional to the charged particle density, and weakly sensitive to their energy. The high-voltage pulse has an amplitude of 340 V for ions (-200 V for electrons), a width of 4 $\mu$s, and a rise time of 60 ns. It is generated by modifying the square pulse generator used in ion beam deflection in a neutron generator \cite{tomic1990}, which uses power FETs to fast switch a HV source.  Figure 2 shows typical ion projection images (2D ion spatial distributions intergrated over the third dimension) with averages of 8 images to increase the signal-to-noise ratio. The units are in pixels, which correspond to approximately 150 $\mu$m.  Figure 2a is a false color plot of an ion image of an UCP at a 20 $\mu$s delay time and initial electron temperature $T_e(0)$ of 100 K, which fits well to a 2D Gaussian profile (figure 2b). The ion images maintain a Gaussian profile during most of the lifetime of the UCP as shown in figure 2c-2f. We note that the ion profiles have a flat top and even a dip at very later times about 150-200 $\mu$s (figs. 2d and 2f, expecially for high $T_e(0)$), and this appears earlier for higher $T_e(0)$. It is currently unknown what causes the flat top and dip in the center of the ion images. 
  
\begin{figure}[htbp]
\begin{center}
\epsfig{file=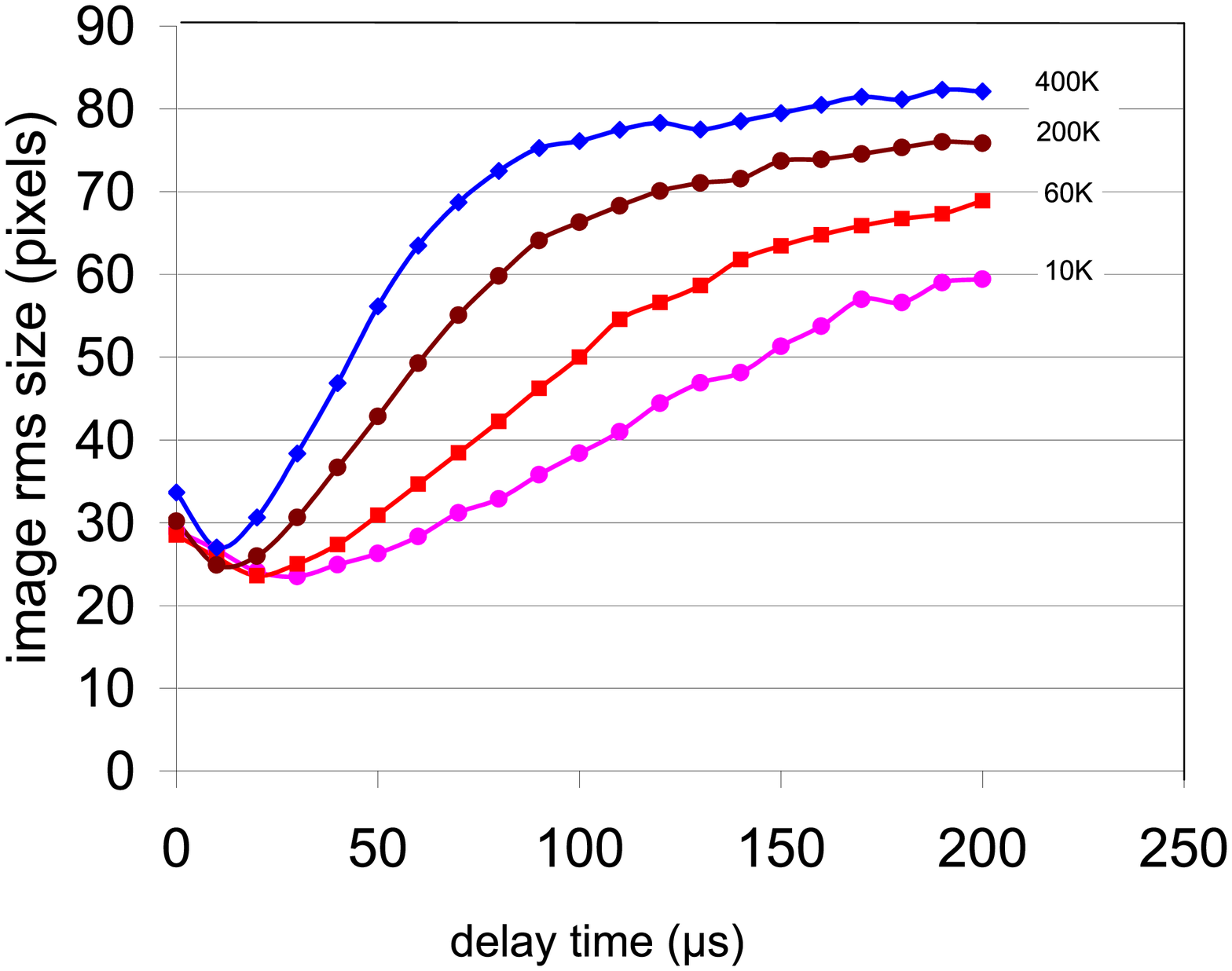, width=5.0in}
\end{center}
\caption{ measured plasma size as a function of elapsed time after the formation of UCP for different initial temperatures. The curves are for $T_e(0)$ of 400 K, 200 K, 60 K and 10 K from top to bottom, respectively. Early in the lifetime of the plasma, the size of the image is dominated by the Coulomb explosion of the dense ion cloud.}
\end{figure}
  
 We extract the plasma size by 2D Gaussian fitting of the ion images (figs. 2c and 2e) at specific delay times after the formation of the UCP. Figure 3 shows the measured plasma sizes as a function of delayed time for different $T_e(0)$. The curves are for $T_e(0)$ of 400 K, 200 K, 60 K and 10 K from top to bottom respectively. Early in the lifetime of the plasma, the measured image size is dominated by the strong Coulomb explosion of the dense ion cloud during transit to the detector, not the true size of the plasma. This is because the electrons are extracted from the UCPs very quickly (a few ns) by the HV pulse, leaving a Gaussian distribution of charged ions. The ions then fly to the detector in about 9 $\mu$s, set by the HV pulse and the accelerating voltages of the other grids (the time-of-flight time of the ions to the detector can be determined from the delay time of the ion current after the formation of the plasma relative to the HV pulse). At early times ($\leq$ 20 $\mu$s), the plasma size is still small (on the order of the intial size, several hundred micros) and the strong Coulomb repulsion between the ions produces a large ion image. As the plasma size increases, the Coulomb explosion effect diminishes and no longer affects the measured size. The image size is at a minimum at about 20 $\mu$s and afterwards increases, reflecting the true size of the plasma with a constant magnification factor of 1.3 (discussed below), as expected from the ballistic expansion model. The size increases slowly and the minimum point of the measured plasma size moves to a later time as we decrease $T_e(0)$, because the expansion velocity which depends on $T_e(0)$ gets smaller, and also the Coulomb explosion effect diminishes more slowly because of the slower expansion.

Assuming that the ion distribution is not affected by the fast HV pulse and maintains a spherical Gaussian distribution during the ions transport to the detector, we can extract the initial ion cloud size from the ion projection image by correcting for the Coulumb explosion effect. This is done as follows: First, we start with the time dependent plasma density distribution $n(r,t)=n_0 (\sigma_0/\sigma_t)^3 e^{-r^2/(2\sigma^2_t)}$; then, we calculate the Coulomb potential of the ion cloud at specific delay time and extract the average acceleration of the ion cloud; next, we obtain the ion cloud size and expansion velocity after the Coulomb explosion with a small time-of-flight step (small enough for constant accerelation for each iteration); finally, we iterate this procedure to get the final ion cloud size after the total time-of-flight, which agrees with the measured ion size. that is, the plasma size indeed follows the ballistic expansion as expected from a simple hydrodymics model througout the whole lifetime of the UCP. This only results in a few percent change in the plasma expansion velocity by including the corrected plasma sizes of the early times compared to that found by only fitting the linear region of the ion image sizes at later times. If we also consider that the ion cloud will freely expand with the ion acoustic velocity in addition to the Coulumb explosion during the time-of-flight, we need to shift the plasma sizes up by several hundred micrometers, which is equivalent to shifting the x-axis (time) in figure 3, but this does not affect the extracted plasma expansion velocity.  At later times, especially for high $T_e(0)$, the measured size does not linearly increase. This is partly because the size of the UCP is large enough to be affected by the 4 posts that secure the grids above and below the plasma (the top and bottom grids), and it approaches the 3-cm size of the detector. 
  
\begin{figure}[htbp]
\begin{center}
\epsfig{file=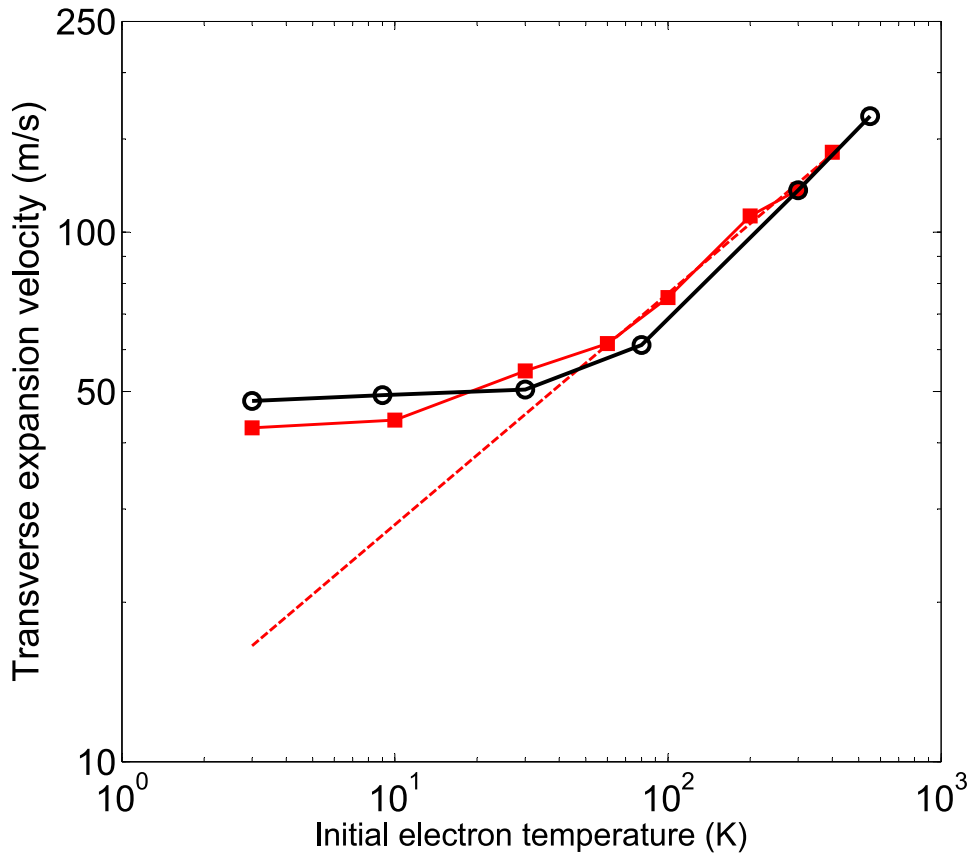, width=5.0 in}
\end{center}
\caption{ The asymptotic expansion velocity as a function of $T_e(0)$. The red solid line with squares is the experimental result which matches the results obtained by measuring the plasma oscillation frequency (the black solid curve with circles)\cite{kulin2000}. The red dashed line is the linear fitting of the data above 60 K with a slope of about 1/2. }
\end{figure} 

By fitting to the sizes after about 20 $\mu$s (for high $T_e(0)$, only fitting the restricted linear region), we can get the asymptotic expansion velocities of UCPs at different $T_e(0)$ with a magnification factor of 1.3 due to an ion lensing effect (the red solid curve with square points in figure 4). The magnification arises from the electric fields which tend to focus or expand the ions (depending on the voltage settings of the grids) as they transport to the detector. It is confirmed by adjusting the voltages on the grids (especially the middle and front grids), which strongly affect the ion image size as well as the scaling factor. By using an ion optics simulation program, we simulate our ion projection imaging setup with the actual spacings and voltage settings of the grids, and find the ion lensingmagnification factor from the trajectories (which is 1.3 for the images at figure 2). At high $T_e(0)$ ($\geq 60$ K), the expansion velocities $v_0 \propto \sqrt{T_e(0)}$ as expected from a simple hydrodymics model, that is, the slope of the red dashed line in figure 4 is about 1/2; at lower $T_e(0)$ ($\leq 60$ K), the expansion velocities are higher than expected from the self-similiar expansion, which indicates heating. The black solid line with open circles is the asymptotic expansion velocity obtained by measuring the plasma oscillation frequency \cite{kulin2000}. The good agreement between our experimental results and earlier results obtained by measuring the plasma oscillation frequency strongly supports the measurement of UCP expansion velocities with both the projection imaging method and the previous technique. 

The excess expansion velocity at low $T_e(0)$ as seen in \cite{kulin2000} and in fig. 4 is attributable to heating due to three-body recombination (TBR) collisions, and was verified in \cite{killian2001}, which directly observed the Rydberg atoms formed in the collisions.The $T_e^{-9/2}$ dependence of TBR makes it important in UCPs, and especially at low $T_e$. 

\begin{figure}[htbp]
\begin{center}
\epsfig{file=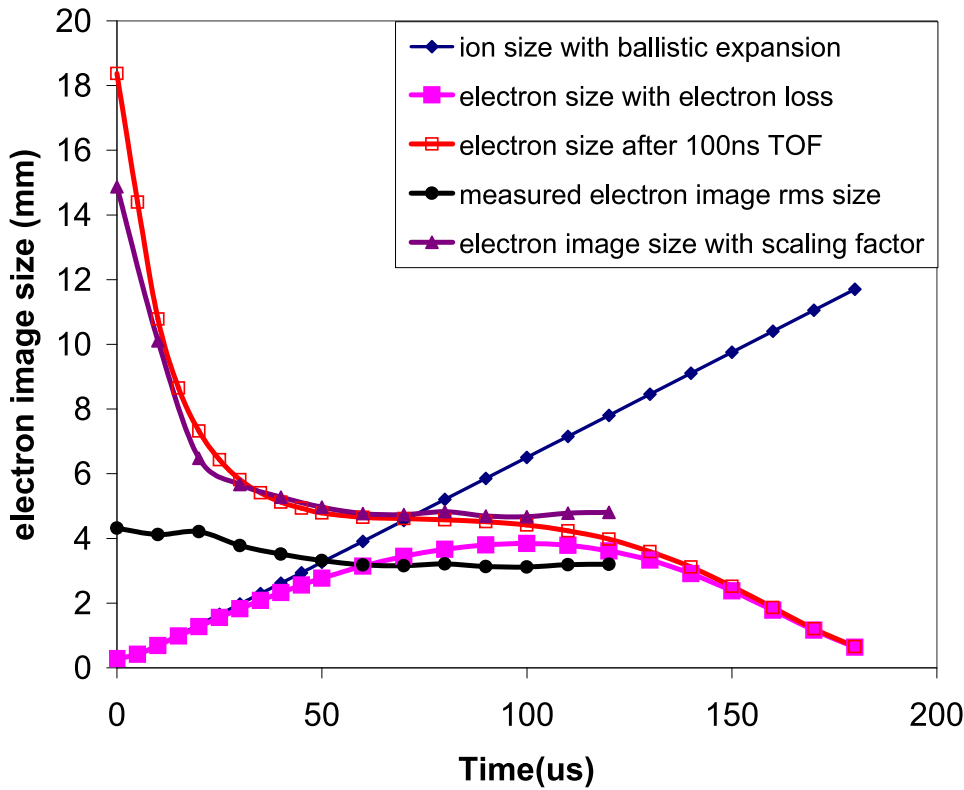, width=5.0in}
\end{center}
\caption{The electron sizes of UCPs as a function of time for $T_e(0)$ = 100 K. The black curve with dots is the electron size extracted from the 2D Gaussian fitting of the electron images. The brown curve with triangles is the electron size with the scaling factor due to the charged particle lensing effect, which is consistent with the theoretical calculation electron size with 100 ns Coulumb explosion time (the red curve with open squares).} 
\end{figure} 

Using the same imaging technique as for ions, we can also image the electrons by reversing the polarity of the HV pulse and the voltages on the grids between the plasma and detector. Figure 5 is the measured electron size as a function of elapsed time after the formation of the UCP. The black curve with circles is the measured electron size extracted from the 2D Gaussian fitting of the images. The brown curve with triangles is the electron size with the actual scaling factor due to an electron lensing effect, which is consistent with the theoretical calculation of electron size with a 100-ns Coulumb explosion time (the red curve with open squares), which also took into account electron loss due to the evaporation of electrons out of the system. We assume that the ion cloud follows a ballistic expansion model (the blue curve with diamonds), while the electron distribution is initially identical to the ion distribution, but with a truncation at the appropriate radius such that the total electron number agrees with the measured charge imbalance at a specific delay time. We then perform self-consistent calculations for the plasma potential to extract the final electron size (the magenta curve with squares).  The electron magnification is obtained from trajectory simulations with the voltage settings of the grids and the ion spatial distribution. We note that the electron lensing factor, unlike that of the ions, is not constant during the whole UCP lifetime due to the strong coulumb force of the ion cloud on the much lighter electrons, especially for the first 30 $\mu$s of the plasma lifetime. The electrons are removed from the plasma in a few ns after applying the HV pulse, but the ions maintain their Gaussian spatial distribution during that short period of time, which exerts a strong Coulomb force on the electrons and partially cancels the applied electric field. This increases the electron lensing, confirmed by the trajectory simulation. As the plasma expands, the plasma size gets larger, and the Coulumb force on the electrons due to the ion cloud gets smaller, so the electron lensing tends towards a constant at later times ($\geq$ 30 $\mu$s). The ion lensing is constant because there are no electrons left during the ion Coulumb explosion phase. 

In conclusion, we have developed a projection imaging technique to study the dynamics of an expanding ultracold plasma. Unlike the previous experimental technique that used the plasma oscillation frequency, which only worked at early times, this method can study expansion dynamics for the entire plasma lifetime. In addition, we can measure both the evolving electron and ion spatial distributions. This method will be usefull for further study of ultracold plasmas, such as plasma instabilities, plasma expansion under different condition of magnetic confinement.

\begin{acknowledgments}
This work was supported by the National Science Foundation PHY-0714381.
\end{acknowledgments}


\end{document}